\documentstyle[12pt]{article}

\begin{document}

\begin{titlepage}
\begin{flushright}
{\large  CINVESTAV-FIS-98-30} 
\end{flushright}
\vskip1.5cm
\begin{center}

{\Large \bf Mixing of two-level unstable systems}
\vskip 2cm

{\large G. L\'opez Castro$^{a}$, J. H. Mu\~noz$^{a,b}$ and J.
Pestieau$^c$} \\

$^a$ {\it Departamento  F\'\i sica, Centro de Investigaci\'on y de
Estudios} \\ {\it Avanzados del IPN, Apdo. Postal 14-740, 07000 M\'exico,
D.F., M\'exico}
\

$^b$ {\it Departamento de F\'\i sica, Universidad del Tolima, }\\
{\it A. A. 546, Ibagu\'e, Colombia}
\

$^c$ {\it Institut de Physique Th\'eorique, Universit\'e Catholique} \\
{\it de Louvain, 2 Ch. du Cyclotron, 1348 Louvain-la-Neuve, Belgium}
 \end{center}
\vskip1.5cm
\begin{abstract}
Unstable particles can be consistently described in the framework of
quantum field theory. Starting from the full S-matrix amplitudes of
$B^+ \rightarrow (2\pi,\ 3\pi) l^+ \nu$ decays as examples in the energy
region where the $\rho-\omega$ resonances are dominating, we propose a
  prescription for the mixing of two quasi `physical' unstable states that
differs from the one obtained from the
diagonalization of the $M-i\Gamma/2$ non-hermitian hamiltonian. We discuss
some
important consequences for CP violation in the $K_L-K_S$ system.

\end{abstract}


\end{titlepage}%

\medskip

\

  The $\rho-\omega$ and $K_L-K_S$ mesons are two beautiful examples of
two-level mixed systems useful to study important properties  of quantum
mechanics and fundamental interactions of unstable particles. The effects
of isospin breaking in the case of $\rho-\omega$ system and CP violation
in the case of $K_L-K_S$ system convert the corresponding 
 eigenstates ($\rho^I, \omega^I$) or ($K_1,K_2$) into physical
eigenstates ($\rho,\omega$) and ($K_L,K_S$). These systems allow to study
the violation of fundamental symmetries where the effects of unstabilities
play an essential role.

  Unstable particles can be consistently treated only in the framework of
quantum field theory\cite{1,2}. They can not
be described by asymptotic states entering the calculation of physical 
S-matrix amplitudes. Instead, they are associated to propagation
amplitudes (propagators) between their production and decay locations and
can not be detached from these mechanisms in order to extract truncated
amplitudes. In quantum field theory, unstable states or resonances are 
special cases of non-perturbative phenomena obtained from a full
resummation of perturbative bubble graphs \cite{1,2}. In addition, the  
space-time behaviour of the amplitudes for production and decay of resonances  
obey, in extremely good approximation, the celebrated exponential decay
law and the covariance properties
for the time-evolution amplitudes \cite{2}. 

   The conventional quantum mechanical treatment of symmetry breaking in
two-level unstable systems consists in finding the eigenstates that
diagonalize a non-hermitian effective hamiltonian of the form $H =
M-i\Gamma/2$
\cite{3,4}, where $M$ and $\Gamma$ are $2 \times 2$ hermitian matrices
describing the mass and decay properties \cite{5} of the unstable states.
$H$ governs the time evolution of the so-called 
 physical eigenstates which at initial time are given by
\begin{eqnarray}
| X \rangle &=& | X^s \rangle + \epsilon | Y^s \rangle\ , \\
| Y \rangle &=& | Y^s \rangle - \epsilon | X^s \rangle\ , \\
\epsilon &=& \frac{ \langle X^s | H^{SB} | Y^s \rangle }{m_X
-m_Y + \frac{i}{2} (\Gamma_Y -\Gamma_X)}\ .
\end{eqnarray}
Here $| Z^s\rangle$ denotes an interaction eigenstate, $m_Z\
(\Gamma_Z)$ is the mass (decay width) of the unstable state and $\epsilon$
is the mixing parameter due to symmetry breaking. $H^{SB}$ is the
symmetry breaking hamiltonian that mixes the $X^s$ and $Y^s$ states. As it
could be easily checked, the physical states are non-orthogonal which can
be traced back to the non-hermitian character of the hamiltonian.

   The purpose of this paper is to demonstrate that the calculation of the
full S-matrix amplitude for a process involving the production and decay
of mixed resonances, leads to a different mixing prescription for the
unstable quasi `physical states' than the one obtained from the
diagonalization of the effective $M -i\Gamma/2$ hamiltonian. 
 In other words, the inclusion of symmetry breaking
in the evaluation of transition amplitudes involving the
approximation where resonances are described by asymptotic states can be
properly done by using the quasi `physical states' as given
below in Eqs. (4)--(5) and not in Eqs. (1) and (2). The numerical impact
of using both approaches in the
evaluation of symmetry breaking when extracting truncated physical
observables as branching fractions for $B \rightarrow V l \nu$ \cite{6,7},
can be very important.

 To be more specific let us consider the S-matrix amplitudes
of the {\it full} decay processes 
$B^+ \rightarrow (2\pi, 3\pi) l^+\nu_l$, which are
dominated by the intermediate $\rho$ and $\omega$ resonances (this example 
illustrates the main characteristics of a two-level unstable mixed 
system). We
show that a convenient prescription for the physical quantum mechanical
eigenstates should be taken as \cite{7}:
\begin{eqnarray}
| \rho \rangle &=& | \rho^I \rangle + \epsilon' | \omega^I \rangle\ , \\
| \omega \rangle &=& | \omega^I \rangle + \epsilon'' | \rho^I \rangle\ ,
\end{eqnarray}
in order to evaluate the matrix elements of the truncated processes $B^+
\rightarrow ( \rho^0,\ \omega)l^+ \nu_l$ in presence of isospin symmetry
breaking. In the above Eqs. $\epsilon'$ and $\epsilon''$ are given by
\begin{eqnarray}
\epsilon' &=& \frac{ m^2_{\rho \omega}
}{m^2_{\rho}-m^2_{\omega} +im_{\omega}\Gamma_{\omega}}\ , \\
\epsilon'' &=& \frac{ m^2_{\rho \omega}}{m^2_{\omega}-m^2_{\rho}
+im_{\rho}\Gamma_{\rho}}\ ,
\end{eqnarray}
where $m_{\rho \omega}^2 \equiv \langle \omega^I | H^{\Delta I=1} | \rho^I
\rangle $ is the $\rho-\omega$ mixing strength.
This results into sizable numerical differences with respect to Eqs.
(1--3) in the evaluation of isospin symmetry breaking effects as
discussed in Refs. \cite{7}.

   Let us consider the full S-matrix amplitude for the semileptonic 
process $B^+ (p_B) \rightarrow \pi^+(p_1)\pi^-(p_2) l^+(p) \nu_l(p')$,
where 
$p_i$ denotes the corresponding four-momenta (the results for the $3\pi l 
\nu_l$ decay mode are straightforward). Including the contributions of
intermediate isospin eigenstates ($\rho^I,\ \omega^I$) and isospin
breaking effects through $\rho-\omega$ mixing \cite{8}, we obtain  (we
assume that only the $\rho^I$ can couple to the $\pi \pi$ system, {\it
i.e.} we ignore a possible {\it direct} contribution $\omega^I \rightarrow
\pi^+\pi^-$ ): 
\begin{eqnarray}
{\cal M}(B \rightarrow 2\pi l \nu) &=& \frac{G_F V_{ub}}{\sqrt{2}}
l^{\mu} \left \{ {\cal M}_{\mu\alpha}( B^+ \rightarrow \rho^{I*}) ({\cal
P}_{\rho})^{\alpha \beta}(q)  \right. \nonumber \\
&& \left. \ + {\cal M}_{\mu\alpha}( B^+\rightarrow \omega^{I*})( {\cal
P}_{\omega})^{\alpha}_{ \nu}(q)\cdot i m_{\rho\omega}^2 \cdot 
({\cal P}_{\rho})^{\nu \beta}
(q) \right \} i g_{\rho \pi \pi} (p_1-p_2)_{\beta} \ .
\end{eqnarray}
Here $G_F$ is the Fermi constant, $V_{ub}$ is the relevant CKM
matrix element,
$g_{\rho \pi \pi}$ is the $\rho \pi \pi$ coupling, $l^{\mu}$ is the
leptonic current and $q^2 \equiv (p_1+p_2)^2$ is the squared invariant
mass of the $2\pi$ system. The hadronic weak matrix element is given by
 (since we neglect the lepton masses we drop the terms
proportional to $(p+p')_{\mu}$) \cite{9} 
\begin{equation}
{\cal M}_{\mu \alpha}(B \rightarrow V^*) = \frac{2}{\Sigma} \epsilon_{\mu
\alpha \rho \sigma} p_B^{\rho}q^{\sigma} V(t) + i \{ g_{\mu \alpha} \Sigma
A_1(t) - \frac{Q_{\alpha}}{\Sigma} (p_B+q)_{\mu} A_2(t)\}
\end{equation}
where $\Sigma \equiv m_B+m_V,\ Q=p_B-q$ ($t=Q^2$) and $V(t),\ A_i(t)$ 
are Lorentz-invariant form factors. The $^*$ symbol means that the vector
meson is produced off its mass-shell. 

The propagators of the resonances are given by:
\begin{equation}
({\cal P}_i)^{\alpha \beta} (q) = \frac {-i  g^{\alpha \beta} }{q^2 -m_i^2
+ im_i\Gamma_i}+ ({\rm terms\ in}\ q^{\alpha}q^{\beta}) .
\end{equation}

  Since the $\rho^I$ coupling to $\pi^+\pi^-$ is a conserved
effective current, {\it i.e.} $q\cdot
(p_1-p_2)=0$, only the transverse component of the vector meson
propagators give a non-zero contribution. In addition, because the
intermediate $\rho^I$ and $\omega^I$ mesons are produced from the
recombination of the daughter $\bar{u}$ (in the $\bar{b} \rightarrow
\bar{u}$ transition) and the spectator $u$ quarks, the
hadronic weak amplitudes are related by 
${\cal M}_{\mu \alpha} (B^+ \rightarrow \omega^I)=
{\cal M}_{\mu \alpha} (B^+ \rightarrow \rho^I)$. Thus, Eq. (8) can be
written as:
\begin{eqnarray}
{\cal M}(B^+ \rightarrow 2\pi l \nu) &=& i\frac{G_F V_{ub}}{\sqrt{2}}
l^{\mu}
{\cal M}_{\mu \alpha}(B^+ \rightarrow \rho^{I*}) \cdot \frac{g^{\alpha
\beta}}{q^2 -m_{\rho}^2 + im_{\rho}\Gamma_{\rho}} \nonumber \\
&& \ \ \times \left \{ 1 + \frac{m_{\rho \omega}^2}{q^2 -m_{\omega}^2 + i
m_{\omega} \Gamma_{\omega}} \right \} \cdot i g_{\rho \pi \pi} 
(p_1-p_2)_{\beta}\ . 
\end{eqnarray}

  A straightforward computation of the $2\pi$ invariant mass distribution
leads to
\begin{equation}
\frac{d\Gamma(B^+ \rightarrow 2\pi l \nu)}{dq^2} =
\frac{\sqrt{q^2}}{\pi} \frac{ \Gamma( B^+
\rightarrow \rho^I(q^2) l^+ \nu)\cdot \Gamma (\rho^I(q^2) \rightarrow
\pi^+\pi^-)}{ (q^2-m_{\rho}^2)^2 + m_{\rho}^2\Gamma^2_{\rho}} \left |
1 + \frac{m_{\rho \omega}^2}{q^2-m^2_{\omega}+im_{\omega}\Gamma_{\omega}}
\right |^2 .
\end{equation}
The $q^2$ in the argument of $\rho^I$ means that decay widths must be
taken with the $\rho^I$ off its mass-shell.

   A very similar evaluation of the $3\pi$ mass distribution in the decay
$B^+ \rightarrow \pi^+\pi^-\pi^0 l^+ \nu_l$ gives (in this case $q^2 =
(p_1 +p_2 +p_3)^2$ corresponds to the $3 \pi$ invariant mass):
\begin{equation}
\frac{d\Gamma(B^+ \rightarrow 3\pi l \nu)}{dq^2} =
\frac{\sqrt{q^2}}{\pi} \frac{ \Gamma( B^+
\rightarrow \omega^I(q^2) l^+ \nu)\cdot \Gamma (\omega^I(q^2) \rightarrow
\pi^+\pi^-\pi^0)}{ (q^2-m_{\omega}^2)^2 + m_{\omega}^2\Gamma^2_{\omega}}
\left | 1 + \frac{m_{\rho 
\omega}^2}{q^2-m^2_{\rho}+im_{\rho}\Gamma_{\rho}}
\right |^2 .
\end{equation}
  The factorization of the decay widths in Eqs. (12) and (13) is an exact
result that follows from the conserved effective current conditions in the
$\rho \rightarrow 2\pi$ and $\omega \rightarrow 3\pi$ vertices.

   The quasi `physical' on-shell decay widths of the $B^+ \rightarrow \rho
l^+
\nu$ and $B^+ \rightarrow \omega l^+ \nu$ decays are obtained by fixing
the $2\pi$ and $3\pi$ invariant masses, respectively,  at the $\rho$
and $\omega$ meson masses (in practice, the 
cuts $m_V^2 -\Delta < q^2 < m_V^2 +\Delta$ are necessary to isolate the
vector mesons from the $q^2$ distribution). Under these conditions we
get:
\begin{eqnarray}
\left. \frac{d\Gamma (B^+ \rightarrow 2\pi l \nu)}{dq^2} \right|_{q^2
 =m_{\rho}^2} &=& \frac{1}{\pi m_{\rho}\Gamma_{\rho}} \Gamma(B^+
\rightarrow \rho^I l \nu) \cdot 
B(\rho^I \rightarrow 2\pi) | 1+\epsilon'|^2, \\
\left. \frac{d\Gamma (B^+ \rightarrow 3\pi l \nu)}{dq^2} \right|_{q^2
=m_{\omega}^2} &=& \frac{1}{\pi m_{\omega}\Gamma_{\omega}} \Gamma (B^+
\rightarrow \omega^I l \nu) \cdot 
B(\omega^I \rightarrow 3\pi) | 1+\epsilon''|^2 \ .
\end{eqnarray}

Therefore, as already pointed out in Ref. \cite{7}, the isospin breaking
effects trough $\epsilon',\ \epsilon''$ must be removed from the measured  
invariant mass distributions quoted in \cite{6} in order to compare
quantities related by isospin symmetry.
   The results given in Eqs. (14) and (15) are identical to the ones
obtained in Ref. \cite{7} where it was assumed that the physical quantum
mechanical eigenstates for the $\rho^0$ and $\omega$ mesons are given by
Eqs. (4)--(5).

  Another way to compare the symmetry breaking effects from the
prescriptions of Eqs. (1--3) and (4--5) is to decompose the resonant
pieces of the amplitudes for $2\pi$ and $3\pi$ semileptonic $B$ decays.
This gives, respectively:
\begin{eqnarray}
\frac{1}{s_{\rho}} \left\{ 1+\frac{m_{\rho \omega}^2}{s_{\omega}} \right
\}&=&
\frac{1}{s_{\rho}} \left\{ 1+\frac{m_{\rho \omega}^2}{\delta^2} \right
\}-\frac{1}{s_{\omega}} \cdot \frac{m_{\rho \omega}^2}{\delta^2}, \\
\frac{1}{s_{\omega}} \left\{ 1+\frac{m_{\rho \omega}^2}{s_{\rho}} \right
\}&=&
\frac{1}{s_{\omega}} \left\{ 1-\frac{m_{\rho \omega}^2}{\delta^2} \right
\}+\frac{1}{s_{\rho}} \cdot \frac{m_{\rho \omega}^2}{\delta^2},
\end{eqnarray}
where $s_V \equiv q^2-m_V^2+im_V\Gamma_V$ and $\delta^2\equiv
m_{\rho}^2-m_{\omega}^2+i(m_{\omega}\Gamma_{\omega}-m_{\rho}
\Gamma_{\rho}) \approx 2\bar{m}\{m_{\rho}-m_{\omega} + i(\Gamma_{\omega}
-\Gamma_{\rho})/2\}$ and $\bar{m}$ is the average mass of $\rho$ and
$\omega$ mesons. Note that the first term in the r.h.s. of Eqs.
(16)--(17) would correspond to the use of Eqs. (1--3) and give equal
strengths for isospin breaking in the $B^+ \rightarrow (\rho^0,\ \omega)
l^+\nu$ decay rates. However, the second terms in the r.h.s. of Eqs.
(16--17) give very different contributions due to the propagation of the
$\omega \ (\rho)$ meson in the $2\pi \ (3\pi)$ channel.

   From Eqs. (14)--(15), the effects of isospin breaking in $B^+
\rightarrow \rho^0 l^+ \nu$ result more important than in the $B^+
\rightarrow \omega l^+\nu$ transition
(because $|1+\epsilon'| \approx 1.18,\ |1+\epsilon''| \approx 1.0$).
This fact is somehow accidental because $m_{\omega}-m_{\rho} \approx
\Gamma_{\omega}$ and therefore the real and imaginary parts in $\epsilon'$
have almost equal weights. This situation is quite similar in the 
$K_L-K_S$ system where $m_{K_L}-m_{K_S} \approx(
\Gamma_{K_S}-\Gamma_{K_L})/2 \approx \Gamma_{K_S}/2$ and, therefore,
there is not an important numerical difference when computing mixing
effects in $K_L\rightarrow 2\pi$ decays through Eqs. (1)--(3)
or (4)--(5). However, the effects are different in CP violating $K_S
\rightarrow 3\pi$ decays.
 As is well known (see \cite{4,10}), the mixing of states accounts
for the complex phase ($\approx \pi/4$) in the CP violation parameters
$\eta_{+-,00}$ measured in $K_L \rightarrow \pi\pi$ decays. According to
the equivalent prescription as the one for the $\rho-\omega$ system, Eqs.
(4)--(5) would 
imply that the complex phase in CP-violating parameters of $K_S
\rightarrow 3\pi$ decays
should be almost zero , which is in
clear disagreement with the results obtained using the conventional
quantum mechanical eigenstates of Eqs. (1)--(3) that predict the same
phase as in $K_L \rightarrow 2\pi$. 

  In practice however, it is difficult to test the difference between both
approaches as far as $B^+ \rightarrow \omega l \nu$ and $K_S \rightarrow
3\pi$ are concerned. .
On the one hand, CP violation (and therefore the complex phase of
$\eta_{+-0,000}$) has not
been observed yet in $K_S \rightarrow 3\pi$ decays so as to test whether
the mixing of unstable states is given by Eqs. (1)--(3) or (4)--(5).
A similar unfortunate situation is present in the $\rho-\omega$ system
because the very narrow width of the $\omega$ meson does not allow to show
up the interference effects due to $\rho-\omega$ mixing by a fine scanning
of the $e^+e^- \rightarrow \pi^+\pi^-\pi^0$ cross section in
the  $\rho-\omega$ region as done in $e^+e^- \rightarrow \pi^+\pi^-$
\cite{11}.

  In conclusion, a consistent treatment of unstable particles as provided
by quantum field theory, leads to a different mixing scheme for
quasi-physical states of a two-level unstable system than the one
obtained from the traditional 
approach based on a $M -i\Gamma/2$ non-hermitian effective hamiltonian.
Symmetry
breaking effects in truncated observables as isospin violation in
semileptonic  $B^+
\rightarrow (\rho^0,\ \omega)$ transitions or CP violation in $K_L-K_S$ 
decays turn out to be very different in both approaches.

\begin{center}
\large Acknowledgements
\end{center}

G.L.C. acknowledges interesting discussions with A. Garc\'\i a. J.H.M.
acknowledges the financial support from Colciencias.


\begin{thebibliography}{99}
\bibitem{5}
A `relativistic' formulation starts from a non-hermitian effective  
hamiltonian given by
$M^2-iM\Gamma$. The denominator in Eq. (3) in this case becomes
$D^r\equiv m_X^2-m^2_Y + i(m_Y\Gamma_Y-m_X\Gamma_X)$ which reduces to $D^r 
\approx 2\bar{m}[m_X -m_Y + i(\Gamma_Y-\Gamma_X)/2] = 2\bar{m} D^{nr}$
when
the mass splitting is negligible as in $\rho-\omega$ and $K_L-K_S$
systems ($\bar{m}$ is the average mass of $X$ and $Y$).
\end{thebibliography}
\end{document}